\documentclass[fp,twocolumn]{jpsj3}

\usepackage{txfonts}
\usepackage{graphicx}
\usepackage{xcolor}
\usepackage{nicefrac}
\usepackage{xfrac}
\usepackage{ulem}
\usepackage{lineno}
\usepackage{amsmath} 
\usepackage{amssymb}
\usepackage{physics}

\newcommand{\biCa}{Ca$_{3}$Ru$_2$O$_7$}

\newcommand{\Rxx} {{\rho}{_{xx}}}
\newcommand{\Rxy} {{\rho}{_{xy}}}
\newcommand{\Sh} {{\sigma}{_{h}}}
\newcommand{\Se} {{\sigma}{_{e}}}
\newcommand{\m} {{\sigma}}
\newcommand{\Rhole} {R{_{h}}}
\newcommand{\Relectron} {R{_{e}}}

\title{Two-carrier Magnetoresistance: Applications to  Ca$_3$Ru$_2$O$_7$}
    
\author{Lakshmi Das$^1$, Yang Xu$^1$, Tian Shang$^1$, 
Alexander Steppke$^1$, Masafumi Horio$^1$, Jaewon Choi$^1$, Simon J{\"o}hr$^1$, Karin von Arx$^1$, Jasmin Mueller$^1$, Dominik Biscette$^1$, Xiaofu Zhang$^1$, Andreas Schilling$^1$, Veronica Granata$^{3}$, Rosalba Fittipaldi$^{2,3}$, Antonio Vecchione$^{2,3}$ and Johan Chang$^1$ }

\inst{$^1$ Physik-Institut, Universit\"{a}t Z\"{u}rich, Winterthurerstrasse 190, CH-8057 Z\"{u}rich, Switzerland\\
$^2$ CNR-SPIN, I-84084 Fisciano, Salerno, Italy\\
$^3$ Dipartimento di Fisica "E.R.~Caianiello", Universit\`{a} di Salerno, I-84084 Fisciano, Salerno, Italy}

\abst{ Ambipolar transport is a commonly occurring theme in semimetals and semiconductors. Here we present an analytical formulation of the conductivity for a two-band system. Electron and hole carrier densities and their respective conductivities are mapped into a two-dimensional unit-less phase space. Provided that one of the carrier densities is known, the dimensionless phase space can be probed through magnetoresistance measurements. This formulation of the two-band model for conductivity is applied to magnetoresistance experiments on Ca$_3$Ru$_2$O$_7$. While previous such  measurements focused on the low-temperature limit, we cover a broad temperature range and find negative magnetoresistance in an intermediate interval below the electronic transition at 48 K.   The low-temperature  magnetoresistance in Ca$_3$Ru$_2$O$_7$ is consistent with a two-band structure. However, the model fails to describe the full temperature and magnetic field dependence. Negative magnetoresistance found in an intermediate temperature range is, for example, not captured by this model. We thus conclude that the electronic and magnetic structure in this intermediate temperature range render the system beyond the most simple two-band model.}

\begin{document}
\maketitle

\section{Introduction}

The combination of electron and hole charge carriers is a common theme of condensed matter physics~\cite{FauquePRM2018}. It is also essential for many applications such as semiconductor junctions, excitonic solar cells and light emitting diodes. Transport properties of semimetals are defined by electron and hole charge carriers. In its simplest form, a semimetal consists of just one electron- and one hole-like band crossing the Fermi level. Exposed to a magnetic field, the electrons and holes experience a Lorentz force which results in a negative or positive Hall effect depending on whether conductivity is dominated by electrons or holes.  Compensated ambipolar transport properties stem from the detailed balance between electron and hole mobilities~\cite{BelPRL2003}. Most semimetals/metals have more than two bands crossing the Fermi level. Nevertheless, a two-band model can still be of some relevance and is widely used to discuss transport experiments~\cite{LIPRL2007,RourkePRB2010,Rana2018,EguchiPRB2019,watts2000evidence,tokumoto2004electric,peramaiyan2018anisotropic,PhysRevB.96.121107,FournierPRB1997}. In some multiband systems, only two Fermi surface sheets host high-mobility carriers -- making the two-band model approximately viable. Orbital selectivity found in systems with strong Hunds coupling~\cite{Georges2013} may also trigger selective Fermi surface gapping while other sheets are untouched -- thus providing another route to two-band physics. Finally, electronic reconstruction of single band systems is a common motif that can generate two-band structures~\cite{HePNAS2019,SachdevJPCM2012,ChangNJP2008}. 
The wide applicability justifies a detailed analysis of the two-band model.

Here we show that the electron and hole transport properties can be described by two dimensionless parameters $\alpha$ and $\beta$. This ($\alpha$,$\beta$) phase space connects smoothly the ambipolar transport with the mono-carrier dominated regime. For constant filling, the system must move along ($\alpha$,$\beta$) specific contour lines. If the carrier density of -- say the electrons -- is known from angle resolved photoemission spectroscopy (ARPES) ~\cite{DamascelliRMP} or quantum oscillation~\cite{SebastianREVIEW} experiments, the temperature dependent parameters $\alpha$ and $\beta$ can be determined from a magnetotransport experiment. We apply this new formulation of the two-band model to magnetoresistance experiments of the semimetallic phase of \biCa. While previous studies focused entirely on the low-temperature limit ~\cite{kikugawa_ca_2010}, we cover a broad temperature range and find negative magnetoresistance for 30 K $ <T< 48$~K.  For selected temperatures ($T\approx 10$~K), the observed low-field magnetoresistance is captured well by the two-carrier model. However, as a function of temperature (warming) the ($\alpha$,$\beta$) parametrisation is inconsistent with a constant filling. For the case of \biCa, the two-band model fails to explain the full temperature and magnetic field dependence of the magnetoresistance. The most pronounced limitation of the two-band model is its inability to explain the negative magnetoresistance found for 30 K $ <T< 48$~K. We discuss these  discrepancies in terms of a two-stage Fermi surface reconstruction.  Below 48~K, the system is a semimetal but with a Fermi surface including more than two sheets. A second reconstruction across 30~K simplifies the Fermi surface such that it hosts only a single electron- and hole-like pocket. Within this structure reasonable consistency with the two band model is reached at low temperatures.

\begin{figure*}[ht!]
\center{\includegraphics[width=\textwidth]{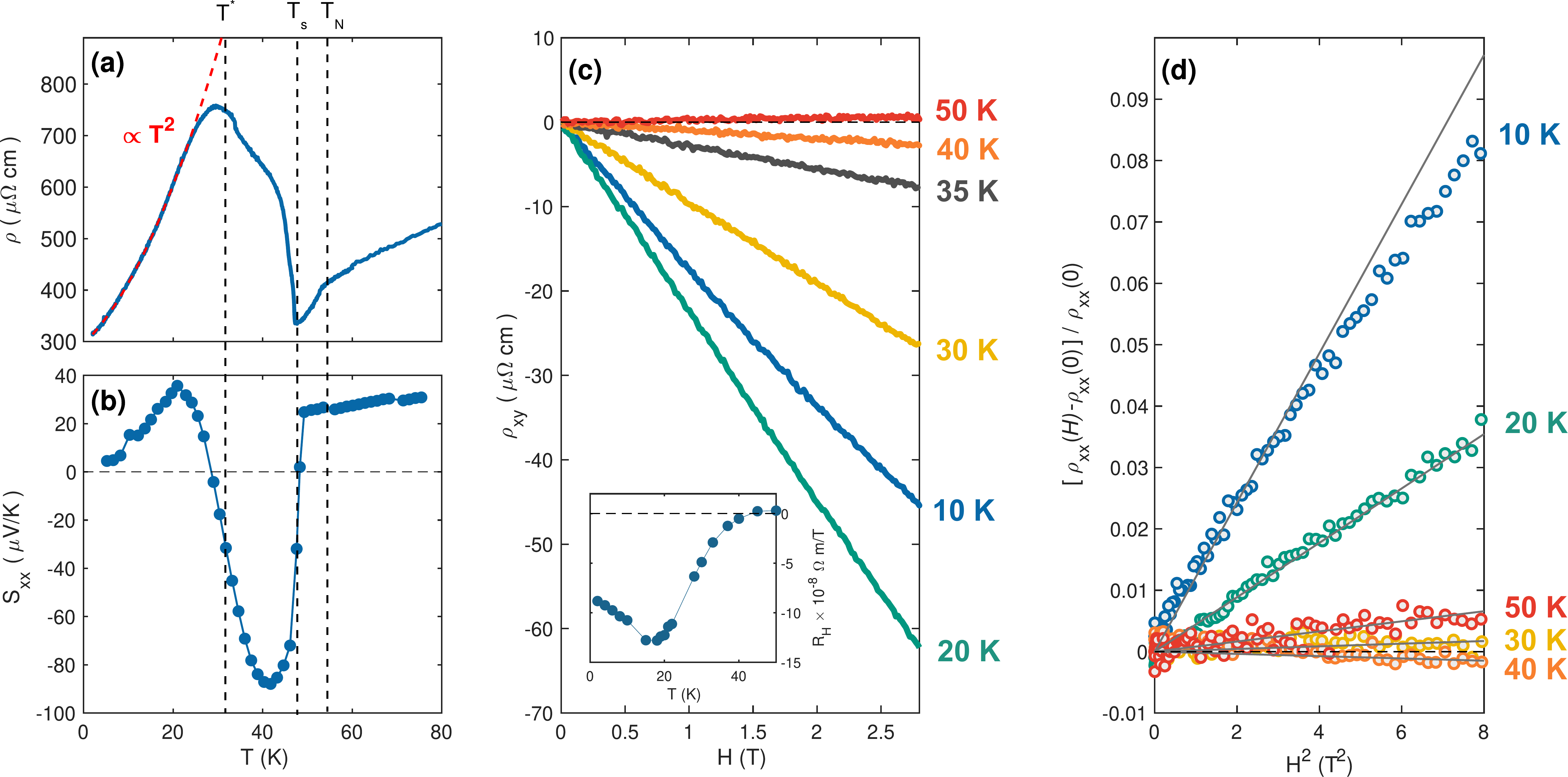}}
\caption{(Color online)  Magnetotransport properties of \biCa. (a,b) Zero-field resistivity and thermopower versus temperature. The vertical dashed lines indicate three temperature scales: $T^*=30$~K, $T_s=48$~K and $T_N=56$~K. These temperature scales represent respectively the onset of the low-temperature metallic transport properties and the AFM-a and AFM-b phases. The $T_s$ temperature scale is also associated with a Fermi surface reconstruction that changes the sign of the thermopower from positive to negative~\cite{XingPRB2018}.  (c) Hall resistivity $\rho_{xy}$ versus magnetic field for temperatures as indicated. The inset shows the temperature evolution of the Hall coefficient, $R_{H}$. (d) Magnetoresistance plotted as $[\rho_{xx}(H)-\rho_{xx}(0)]/\rho_{xx}(0)$ versus magnetic field squared for temperatures as indicated.  Solid grey lines are linear fits.}
\label{fig1}
\end{figure*}

\section{Methods} 

High quality single crystals of \biCa\ were grown by the floating zone techniques~\cite{FukazawaPhysB00,snakatsujiJSSCHEM2001}. The orthorhombic (\textit{Bb2$_1$m}) crystals were detwinned, by pressing along an orthorhombic axis, using a thermo-mechanical detwinning device~\cite{Burkhardt1995} in combination with a polarized light microscope. The longitudinal resistivity (along the $b$-axis), Hall, magnetization and thermoelectric  measurements were carried out in a Quantum Design Magnetic Property Measurement system and a Physical Property Measurement System with magnetic field $H$ applied along the crystallographic $c$-axis. For the thermoelectric experiments, temperature gradients along the orthorhombic $b$-axis (the longer lattice parameter) were recorded using Cernox chips~\cite{Destraznpj2020}.

\section{Results} 
In Figs.~\ref{fig1} a,b, we are plotting the zero magnetic field resistivity $\rho$ (along \textit{b} - axis) and thermopower $S$ versus temperature. The temperature dependence of both curves are in agreement with all previous experiments~\cite{yoshida_quasi-two-dimensional_2004,XingPRB2018,IWATAjmmm2007}. Resistivity displays a weak drop across the AFM-a  N\'{e}el ordering temperature $T_N=56$~K and a sharp upturn below the spin-flip transition $T_s=48$~K into the AFM-b phase. The latter transition is associated with a Fermi surface reconstruction, signaled by a sharp sign change of the Seebeck coefficient~\cite{XingPRB2018} (Fig.~\ref{fig1}b). Below 30~K, the resistivity recovers a metallic Fermi-liquid-like ($\rho\propto T^2$) temperature dependence (Fig.~\ref{fig1}a). Hall resistivity isotherms ($\rho_{xy}$ versus magnetic field $H$) -- shown in Fig.~\ref{fig1}c for temperatures as indicated -- are also consistent with previous reports. Negative Hall resistivity is found for $T<T_s$~\cite{yoshida2007hall}. 

\begin{figure*}[ht!]
\center{\includegraphics[width=1\textwidth]{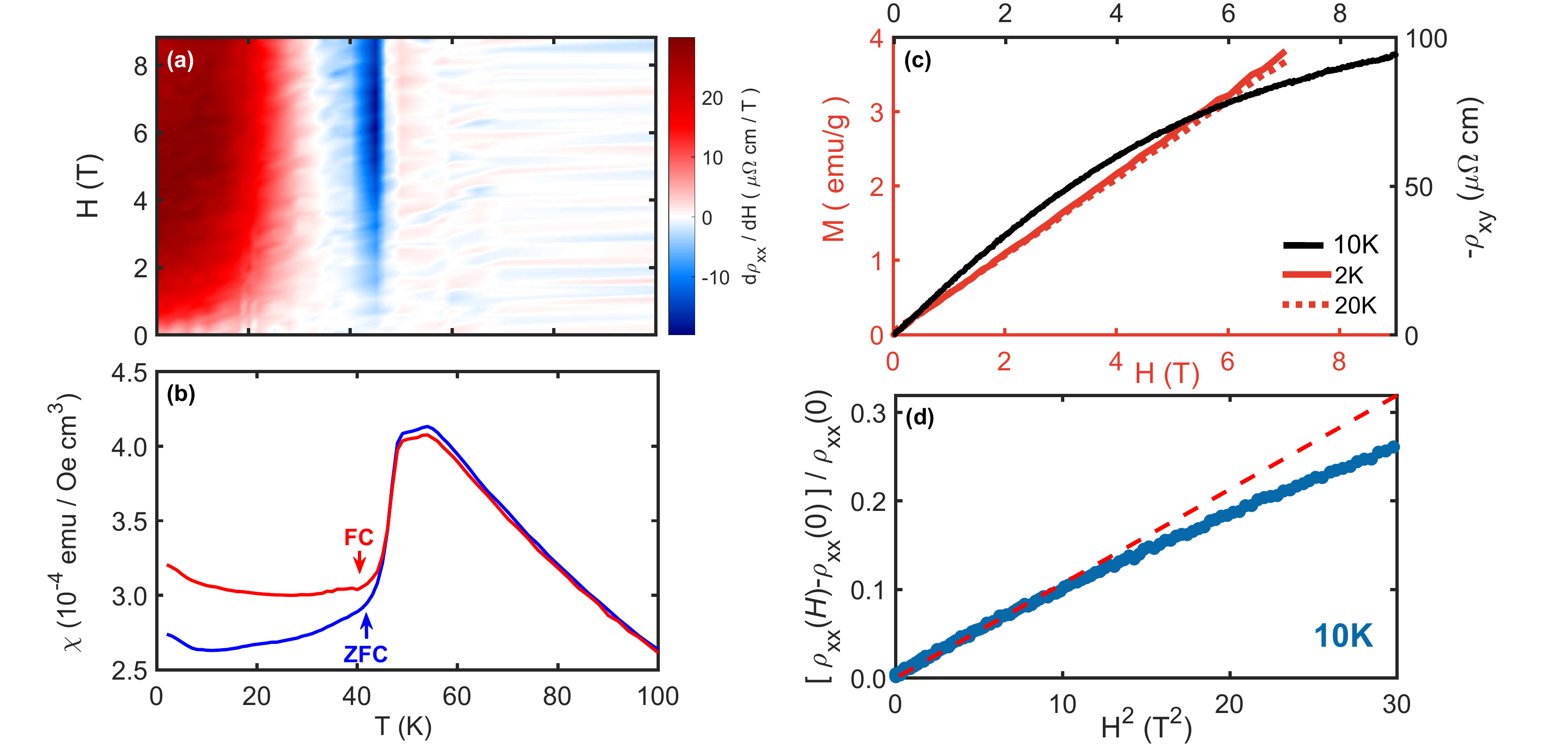}}
\caption{ (Color online) Magnetic field derivative of resistivity d$\Rxx$/dH and magnetic susceptibility. (a) Magnetic field derivative of resistivity, d$\Rxx$/dH plotted in false color scale, as a function of magnetic field and temperature. Red (blue) color indicates positive (negative) magnetoresistance. (b) Susceptibility recorded with an applied field of 0.5~T versus temperature in field-cooled (FC - dotted black) and zero-field-cooled (ZFC - blue curve) mode.  The bifurcation below 48~K demonstrates the existence of a ferromagnetic coupling~\cite{sokolov2019metamagnetic}. (c) and (d) shows the low-temperature magnetization, resistivity and Hall isotherms. (c) Hall and magnetization isotherms plotted as a function of magnetic field $H$ for temperatures as indicated. (d) Resistivity isotherm versus $H^2$. The dashed red line represents the two-carrier model including higher order terms in $H^2$.} 
\label{fig2}
\end{figure*}

Magnetoresistance plotted as $[\rho_{xx}(H)-\rho_{xx}(0)]/\rho_{xx}(0)$ versus $H^2$ is displayed in Fig.~\ref{fig1}d for temperatures as indicated. The low-temperature positive magnetoresistance is consistent with previous studies~\cite{KikugawaJPSJ2010,XingPRB2018}, all of which showed, $[\rho_{xx}(H)-\rho_{xx}(0)] /\rho_{xx}(0)=(\mu_{MR}H)^2$ with mobility $\mu_{MR}\approx0.1$~T$^{-1}$. Here we focus on the  temperature dependent magnetoresistance in Ca$_3$Ru$_2$O$_7$. For $T>T_s$, a small positive magnetoresistance is found whereas for 30~K$<T<T_s$ it is negative before turning to large positive values in the $T\rightarrow0$ limit. The derivative $d\rho_{xx}/dH$ is shown in Fig.~\ref{fig2}a as a function of temperature and magnetic field. The onset of negative magnetoresistance and history dependent magnetic susceptibility (Fig.~\ref{fig2}b) clearly correlate. Both effects appear below $T_s$.

The low-temperature magnetization and Hall resistivity isotherms display different field dependence (Fig.~\ref{fig2}c). The magnetization $M$ is linear in magnetic field  whereas $\rho_{xy}$ is displaying non-linear field dependence suggesting that these two quantities are not directly coupled. The linear trend of the low temperature magnetization data (Fig.~\ref{fig:70028Fig2}c) excludes the possibility of an anomalous Hall effect. We also point out that  although $\rho_{xx}\propto H^2$  in the low-field ($H\rightarrow0$) regime, significant deviations are observed in larger magnetic field (see Fig.~\ref{fig2}d and Reference.\cite{KikugawaJPSJ2010}). In what follows, we discuss the low-temperature large positive and intermediate-temperature negative magnetoresistance behaviour within the two-band model.

\section{Modelling}

 For metals with two Fermi surface sheets of different band curvature, the magnetoresistance is given by~\cite{Ashcroft,LIPRL2007,RourkePRB2010,Rana2018,EguchiPRB2019}: 
 \begin{equation}
\Rxx(H) = \frac{(\Sh+\Se)+\Sh\Se(\Sh\Rhole^2+\Se\lvert{\Relectron}\rvert^2)H^2}{( \Sh+\Se)^2+ \Sh^2\Se^2(\Rhole-\lvert\Relectron\rvert)^2H^2 }~~,
\end{equation}

\begin{equation}
\frac{\Rxy(H)}{H}= \frac{\Sh^2\Rhole-\Se^2\lvert\Relectron\rvert-\Sh^2\Se^2\Rhole\lvert\Relectron\rvert(\Rhole-\lvert\Relectron\rvert)H^2}{( \Sh+\Se)^2+ \Sh^2\Se^2(\Rhole-\lvert\Relectron\rvert)^2H^2 }~~.
\end{equation}

The model consists of four parameters $\sigma_e,\sigma_h$, $R_e$ and $R_h$. The hole and electron conductivities are denoted by ($\sigma_h, \sigma_e$) and the hole and electron Hall coefficients are indicated by ($R_h, R_e$). Given that $\sigma_i=n_i e \mu_i$ and $R_i=\pm 1/(n_i e)$ where $i={e,h}$, the model can also be expressed in terms of electron  and hole carrier density ($n_e$, $n_h$) and mobility ($\mu_e$, $\mu_h$). This model generally assumes that the mobilities are independent of the applied magnetic field. Analysis of $\rho_{xx}$ and $\rho_{xy}$ isotherms  in terms of the four parameters ($\sigma_e,\sigma_h$,$R_e$,$R_h$) is typically associated with ambiguity~\cite{EguchiPRB2019}. Although $\sigma_e$ and $\sigma_h$ are coupled parameters, since in zero-field, $\rho_{xx}=(\sigma_e+\sigma_h)^{-1} = \sigma^{-1}$, the problem has three free parameters.

\begin{figure*}[ht!]
\center{\includegraphics[width=1\textwidth]{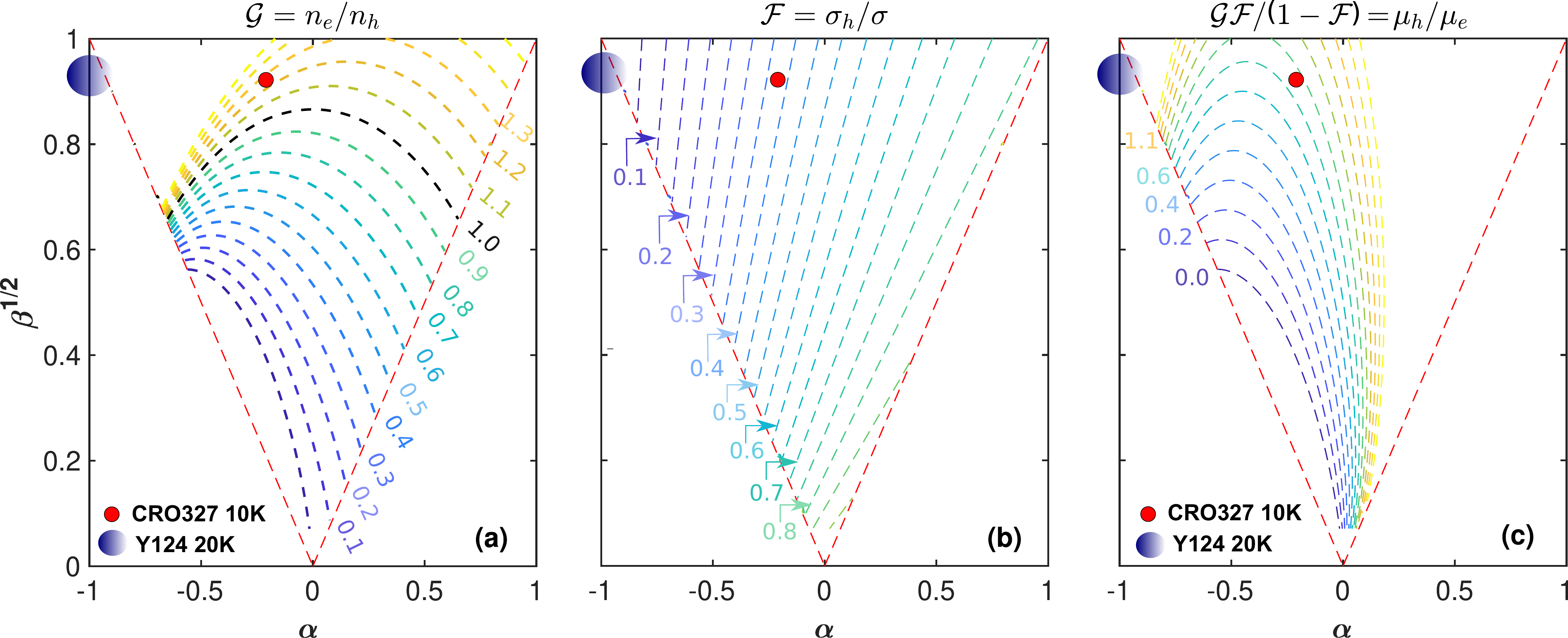}}
\caption{(Color online) Two-band model solutions. (a-c) Carrier density ratio $\mathcal{G}=n_e/n_h$, conductivity ratio $\mathcal{F}=\sigma_h/\sigma$, and mobility ratio $\mu_h/\mu_e$ versus the dimensionless parameters $\alpha$ and $\sqrt{\beta}$. Dashed lines are contours of the respective quantities. Solutions are found for $\beta>\alpha^2$ (lower bound is red line). The $n_e/n_h=1$ contour line, given by $\beta=3(1-\alpha^2)/4$  indicates the border between hole and electron "doped" regimes. The parameter space around $\beta=\alpha=1$ represents the situation $n_e>n_h$ and $\mu_h\gg\mu_e$ whereas $\beta=-\alpha=1$ describes  $\mu_h>\mu_e$ and $n_e\gg n_h$. Hole rich ($n_h\gg n_e$) solutions with either large hole or electron mobility is found in the $\beta,\alpha \rightarrow 0$ limit.  Red and blue circles in (a-c) represents the phase space position of \biCa\ (CRO327) at $T=10$ K and YBa$_2$Cu$_4$O$_8$ (Y124) at $T=20$ K~\cite{RourkePRB2010}~ respectively. } 
\label{fig3}
\end{figure*}

To make progress, we assume that the electron carrier density $n_e$ is known, for example, from ARPES~\cite{DamascelliRMP} or quantum oscillation~\cite{SebastianREVIEW} experiments. The problem then reduces to two equations with two unknowns ($\sigma_h$ and $n_h$), lifting the ambiguity. In the low-field  $H\rightarrow0$ limit, the problem can be expressed in terms of two dimensionless parameters $\alpha$ and $\beta$: 

\begin{equation}
    \alpha=\frac{R_H}{\lvert{R_e}\rvert}=\frac{\rho_{xy}}{H \lvert{R_e}\rvert}=\frac{\Sh^2\Rhole/\lvert{\Relectron}\rvert-\Se^2}{\sigma^2}~~,
\end{equation}
and 
\begin{equation}
    \beta=\frac{3\mathcal{C}}{\sigma \lvert{R_e}\rvert^2}=3\Sh\Se\frac{\left[\Sh\Rhole/\lvert{\Relectron}\rvert+\Se \right]^2}{\sigma^4}~~.
\end{equation}

with $\mathcal{C}$ defined by $\rho_{xx}=\rho_0+\mathcal{C}H^2$ where $\rho_0$ is the residual resistivity. Solving with respect to $\sigma_h$ and $R_h$ yields $\sigma_h/\sigma=\mathcal{F}(\alpha,\beta)$, $R_h/\Relectron$=  $\mathcal{G}(\alpha,\beta)=[\alpha+(1-\mathcal{F})^2]/\mathcal{F}^2$  and  $\mu_h/\mu_e=\mathcal{G}\mathcal{F}/(1-\mathcal{F})$ where the analytical function $\mathcal{F}$ is expressed in the appendix. These three ratios ($n_e/n_h$, $\sigma_h/\sigma$ and $\mu_h/\mu_e$) are plotted in Fig.~\ref{fig3} within the "phase space" of the two dimensionless parameters $\alpha$ and $\beta$. 

These parameters are measurable (modulus $R_e$) and typically temperature dependent. Perhaps $\alpha=R_H/\lvert{R_e}\rvert$ is the most intuitive parameter. Within an electron-hole two-band model, it varies in the interval $[-1,\infty]$. Ambipolar transport, defined by comparable electron and hole contributions to the conductivity, typically leads to  $\vert\alpha\vert \sim 0$. By contrast, for $\alpha \approx -1$ conductivity is dominated by electrons. The second dimensionless parameter $\beta$ is given by the ratio between the magnetoresistance coefficient $\mathcal{C}$ and the conductivity modulus $R_e^2$. 

We notice that the two-carrier model was recently expressed in a similar scheme with two dimensionless parameters~\cite{EguchiPRB2019}. In the work of Eguchi and Paschen (Ref.~\citenum{EguchiPRB2019}),  magnetoresistance is expressed in terms of dimensionless carrier and mobility ratios. Here, we use the inverse approach where the
carrier and mobility ratios are expressed in terms of dimensionless parameters $\alpha$ and $\beta$ obtained from magnetoresistance. This formulation has some advantages. With our notation,  solutions  have the following properties. First, for $\mathcal{F}$ to take real values, $\beta>\alpha^2$ and hence $C/(\sigma R_H^2)>{1/3}$ must be satisfied.
As  $C$, $\sigma$ and $R_H$ are all measurable quantities, this provides a test as to whether a two-band model is applicable. We notice that this is independent of our initial input
for $R_e$. Second, constant filling implies that $n_e+n_h=n_e(1+\mathcal{G}^{-1})$ is invariant. Therefore, if $n_e$ remains constant, the system must stay on a $\mathcal{G}$-contour.
These contour lines "flow" from the electron to the hole dominated limit via the ambipolar~\cite{BelPRL2003} regime (see Fig.~\ref{fig3}a).

For example, increasingly negative Hall coefficient implies $\alpha\rightarrow -1$ and hence $\beta$ should decrease for the system to stay on a constant $n_e/n_h$ contour. Since both $\sqrt{\beta}$ and $\alpha$ scale with $1/R_e$, the shape of the contour lines are independent of $R_e$. A third solution property stems from the $n_e/n_h=1$ contour line~\cite{EomPRB2012,PisoniPRB2016,RULLIERALBENQUE2016164} given by $\beta=3(1-\alpha^2)/4$. If a system is known to be hole-doped, i.e. $n_h>n_e$ ($\mathcal{G}<1$), it must satisfy $\vert\alpha\vert <\sqrt{3/7}$ and $\beta<3(1-\alpha^2)/4$. Deviation from any of the three properties implies: (a) non-constant filling, (b) the system is not having a two-band structure or (c) the two-band model assumptions are too simplistic.

\section{Discussion}

\begin{figure*}[ht!]
\center{\includegraphics[width=\textwidth]{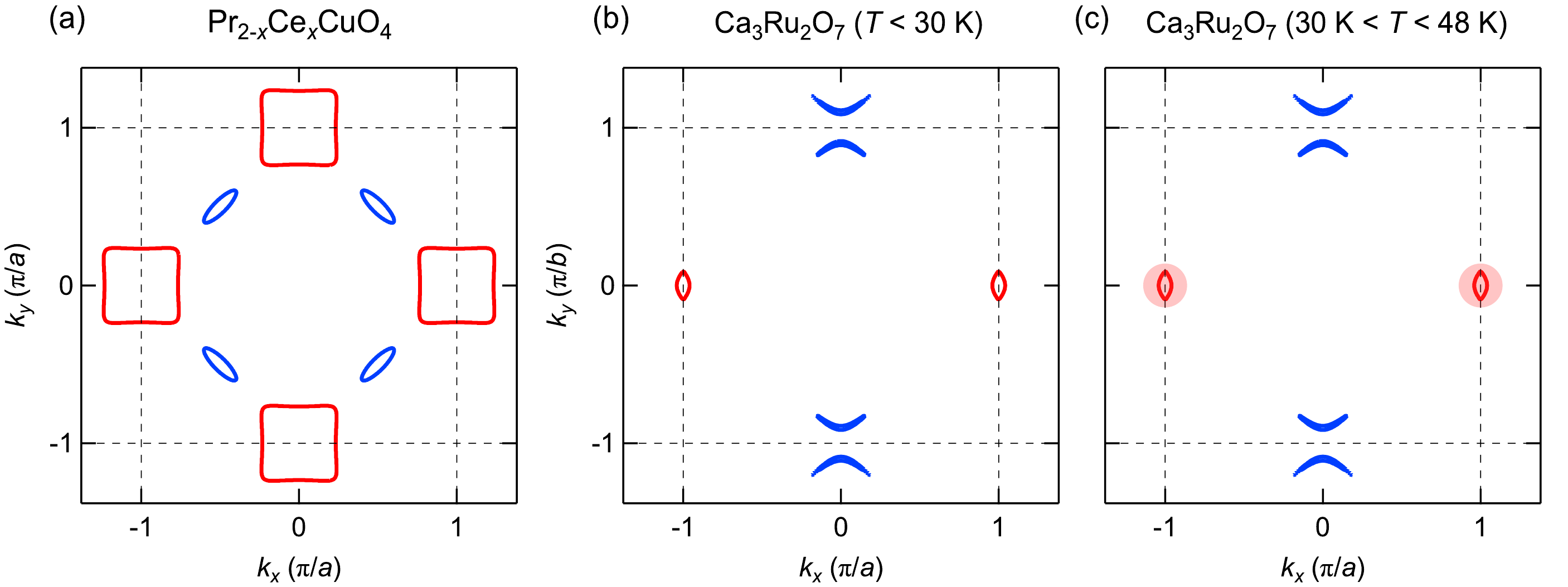}}
\caption{(Color online) Fermi surfaces of two-carrier systems. Blue and red sheets represent respectively hole and electron-like band curvature. (a) The Fermi surface that appears in electron doped cuprates after folding/reconstruction around the anti-ferromagnetic zone boundary.  (b) Low-temperature ($T<30$~K) Fermi surface of \biCa, revealed by ARPES experiments~\cite{Horio2019}. (c) Fermi surface structure of \biCa\ in the  ($30<T<48$~K) temperature interval.  ARPES experiments indicates a more complex structure around the electron pocket. }
\label{fig4}
\end{figure*}

Before analysing the magnetoresistance of \biCa, we start by illustrating the use of the aforementioned solution properties. In doing so, we consider results obtained on hole and electron doped cuprates. For hole doped YBa$_2$Cu$_3$O$_{7-x}$ and YBa$_2$Cu$_4$O$_8$, quantum oscillation experiments yields an electron-like Fermi surface sheet with an area corresponding to $R_e=-29~$mm$^3$C$^{-1}$~\cite{YellandPRL2007,Doiron-Leyraud-Nat2007}. Since the system is hole-doped, we expect $n_h>n_e$. Hall effect experiments suggest that $\alpha\approx -1$~\cite{LebeoufNature2007}. Already here a contradiction emerges as $\vert\alpha\vert >\sqrt{3/7}$ is impossible for a two-carrier hole-doped system. Evaluation of $\beta$ positions the system in the $n_e\gg n_h$ region where $\sigma_h\rightarrow 0$~\cite{RourkePRB2010}. In fact $\beta>\alpha^2$ is not strictly satisfied. A plausible explanation for these contradictory results is that the Fermi surface structure contains more than two sheets. Since the Fermi surface is likely reconstructed by an incommensurate charge-density-wave order~\cite{Wu11,GhiringhelliSCI12,chang12}, a more complex multi-band structure is expected~\cite{MillisPRB2007}. On the electron doped side  ($n_e>n_h$), by contrast, the Fermi surface is expected to fold around the antiferromagnetic zone boundary leading to a two-band structure~\cite{HePNAS2019} (see Fig.~\ref{fig4}a).

Quantum oscillation Hall experiments on Pr$_{1.86}$Ce$_{0.14}$CuO$_4$ (PCCO)~\cite{Breznay} and Nd$_{1.85}$Ce$_{0.15}$CuO$_4$~(NCCO) \cite{HelmPRL2009,HelmPRB2015} reveal a small hole pocket with $n_h\approx 4.0\times10^{25}/m^3$ giving $R_H \ll R_h$. To compare to our model results, we use momentarily $n_e\leftrightarrow n_h$ and $\sigma_e \leftrightarrow \sigma_h$. We therefore have $\vert\alpha\vert\ll 1$ and evaluation of the magnetoresistance for both NCCO and PCCO~\cite{Breznay,GollnikPRB1998} yields $\beta\approx0.03$ less than $3(1-\alpha^2)/4$ as expected. These systems are thus not incompatible with a two-band model.

Next, we discuss our magnetoresistance experiments on \biCa. Recent  ARPES data suggest that \biCa, at base temperature, is a semimetal with a single electron and two identical hole pockets~\cite{Horio2019,markovi2020,PuggioniPRR2020} (see Figure.~\ref{fig4}b).  The electron pocket - corresponds to a carrier density $n_e=7.8\times10^{24} /m^3$, consistent with the dominant quantum oscillation frequency~\cite{KikugawaJPSJ2010}. The electron Hall coefficient $R_e=-1/(e n_e)$ is therefore known  within the confidence provided by ARPES and quantum oscillation experiments. Our low magnetic-field magnetoresistance data permit extraction of $R_H$, $\sigma$ and $\mathcal{C}$  as a function of temperature. At $T=10$~K, we find $\alpha=-0.21$ and $\beta=0.85$ and with that, all model parameters are now known: $\sigma_h=0.37\sigma=0.6\sigma_e$, and $\lvert R_h/R_e \rvert=1.26$ (see Fig.~\ref{fig3}). The electron-hole carrier ratio is consistent with the low-temperature ARPES derived Fermi surface~\cite{Horio2019}. Our results also imply that the electron carriers are more mobile $\mu_e=1.33\mu_h$ than the holes. This is consistent with the fact that the largest quantum oscillation amplitude~\cite{KikugawaJPSJ2010} is found for the frequency corresponding to the electron pocket in the ARPES experiment~\cite{Horio2019,markovi2020}. So far, the two-band model seems to sensibly describe the magnetoresistance experiment. However, there are at least three problematic issues with the two-band model. (i) Even in the temperature range 2 K -- 30 K, $\alpha$ and $\beta$ vary in a fashion that is inconsistent with a constant $n_e/n_h$. (ii) For intermediate temperatures negative magnetoresistance is observed. (iii) We used the low-field limit $H\rightarrow0$ for our analysis. On top of this, higher order magnetic field terms can be derived. However, these higher-order terms are not large enough to account for  the deviation from $\rho_{xx} \propto H^{2}$ under higher fields (Fig. ~\ref{fig2}d).

In what follows, we discuss possible reasons for these discrepancies. We start by stressing that the underlying assumption that carrier mobility is magnetic-field independent is not the sole culprit. Field dependent mobilities could explain (iii) but not (i) and (ii). We also point out that carrier density ratio $n_e/n_h$ may, in fact,  be temperature dependent. Recent density-functional-theory (DFT) calculations suggest a strong coupling between detailed lattice structure and a temperature dependent Coulomb interaction~\cite{PuggioniPRR2020}. Early neutron diffraction experiments indeed demonstrated the temperature dependence of ruthenium-oxygen bond angles~\cite{yoshida_crystal_2005}. It is therefore not inconceivable that the chemical potential and even the low-energy electronic structure is temperature dependent. In fact, recent ARPES studies reported a two stage reconstruction with characteristic 
temperatures of  $T^*=30$ K and $T_s=48$~K~\cite{Horio2019,markovi2020}. Since the negative magnetoresistance is found for $T^*<T<T_s$, it is very likely linked to the intermediately reconstructed electronic structure (see Figure.~\ref{fig4}c). ARPES experiments suggest that these Fermi surface consists for more than two sheets~\cite{Horio2019}. Furthermore, out-of-plane resistivity indicate that a finite inter-layer interaction is still present~\cite{yoshida_quasi-two-dimensional_2004}. As such the electronic structure contains complexity beyond the most simple two-band model.  In addition, our magnetization data (see Figure.~\ref{fig2}b) suggests the existence of weak ferromagnetism (on top of the antiferromagnetic coupling) below $T_s$. It is not uncommon that magnetic correlations  generate a negative magnetoresistance \cite{PhysRevB.66.024433,yamada1972negative} 
that is comparable or larger than the orbital component. It is also not unusual that the ferromagnetic contribution is largest just below the ordering temperature~\cite{Destraznpj2020}. The negative magnetoresistance in \biCa\ therefore is likely linked both to the electronic structure and ferromagnetic properties.

\section{Conclusions}

In summary, we have carried out a magnetoresistance study of Ca$_3$Ru$_2$O$_7$. As a function of temperature, three different regimes are identified: weak positive  magnetoresistance for $T>48$~K, negative magnetoresistance for 30 K $ < T < 48$~K and large positive magnetoresistance below $30$~K. These characteristic temperatures are directly linked to reconstructions of the low-energy electronic structure. We analysed the low-temperature magnetoresistance within a two-band model. An analytical solution to the two-band model for conductivity is developed and expressed in terms of two dimensionless (but measurable) parameters. Whereas reasonable values of electron/hole carrier density and mobility is found for selected temperatures ($T\approx10$~K), the two band model is not capturing the full magnetic field and temperature dependence. The most pronounced limitation of the model is its inability to explain negative magnetoresistance. On top of a two-stage electronic reconstruction, we argue that the Fermi surface structure -- and with that the electron/hole carrier density -- is temperature dependent. This in combination with ferromagnetism generates the complicated magnetoresistance in Ca$_3$Ru$_2$O$_7$. \\[2mm] 

\begin{acknowledgements}

We thank Benoit Fauque for fruitful discussions. L.D., Y. X, M.H., K.v.A, T.S., and J.C. acknowledge support by the Swiss National Science Foundation. L. D. was partially funded through Excellence Scholarship by the Swiss Government.

\end{acknowledgements}

\appendix
\section{}
In the main text, the dimensionless parameters $\alpha$ and $\beta$ are expressed in terms of $\Sh/\m =~ \mathcal{F}(\alpha,\beta)$ and $\Rhole/\Relectron=\mathcal{G}(\alpha,\beta)$ -- Equations (3) and (4). Solving with respect to   $\mathcal{F}(\alpha,\beta)$ and $\mathcal{G}(\alpha,\beta)$ yields:

\begin{equation}
    \mathcal{F}(\alpha,\beta)=
      A_1 +\displaystyle{\frac{A_2}{(A_3+A_4)^{1/3}}} + \frac{2^{2/3}}{6}\left(A_3+A_4\right)^{1/3} ,
\end{equation}

and 

\begin{equation}
    \mathcal{G}(\alpha,\beta)=\frac{{\alpha}+[1-\mathcal{F}(\alpha,\beta)]^2}{[\mathcal{F}(\alpha,\beta)]^2}~~.
\end{equation}

with
 
\begin{equation}
\label{eqn:eqlabel}
\begin{aligned}
	A_1 &= 1+\frac{2}{3}{\alpha},\\
	A_2 &=  \frac{\sqrt[3]{2}}{3}\left({\alpha^2}-\beta\right),\\
	A_3 &= -[ 2{\alpha^3}+ 3\beta(2\alpha+3)],\\
	A_4 &= \sqrt{3}\beta \sqrt{\frac{12\alpha^3(1+\alpha)}{\beta}+(27+36\alpha+8\alpha^2)+\frac{4\beta}{3}}~~.
\end{aligned}
\end{equation}

It is now possible to express the longitudinal and transverse resistance in terms of $\mathcal{F}$ and $\mathcal{G}$:

\begin{equation}\label{eq:rhoxxfull}
    \rho_{xx}=\frac{1}{\sigma}\frac{1+\mathcal{F}(1+\frac{\mathcal{F}\mathcal{G}^2}{1-\mathcal{F}})(\mu_e H)^2}{1+(\mathcal{F}(1-\mathcal{G})\mu_e H)^2} ,
\end{equation}

\begin{equation}\label{eq:rhoxyfull}
    \rho_{xy}=\mathcal{F}\mathcal{G}\frac{1-\left(\frac{1-\mathcal{F}}{\sqrt{\mathcal{G}F}}\right)^2+(1-\mathcal{G})(\mu_e H)^2}{1+(\mathcal{F}(1-\mathcal{G})\mu_e H)^2} R_e H .
\end{equation}
\\
For a semimetal with $\mathcal{G}=1$ ($n_e=n_h$), Equations~\ref{eq:rhoxxfull} and~\ref{eq:rhoxyfull}  simplify to $\Delta\rho(H)/\rho(0)=\mu_e\mu_h B^2$ and $\rho_{xy}=(1-\mu_e/\mu_h)R_e H$. Therefore, no saturation of magnetoresistance or Hall effect is expected. For $\mathcal{G}\neq 1$, however, the magnetoresistance should eventually saturate and non-linear $\rho_{xy}$ isotherms should occur in the high-field limit.  As shown in Fig.~\ref{fig2}d, we indeed observe saturation of the $\Delta\rho(H)/\rho(0)\propto H^2$ and $\rho_{xy}$ as magnetic field strength is increased. In the same field range, the magnetization scales perfectly with $H$. Hence ferromagnetism as a source for the saturating magnetoresistance can thus be excluded. As we find $\mathcal{G}$ close to unity, this saturation effect is stronger than what can be explained by the two-carrier model. It is possible to fit the entire field range, however the resulting parameters ($\sigma_h$,$\sigma_e$,$n_h$,$n_e$) are difficult to reconcile with the quantum oscillation and ARPES experiments. We are therefore concluding that the electron mobility is magnetic field dependent for $H>4$~T.

\end{document}